# Harmonic lasing of x-ray free electron laser: on the way to smaller and cheaper[*]

DENG Hai-Xiao(邓海啸)[1)]  DAI Zhi-Min(戴志敏)

Shanghai Institute of Applied Physics, the Chinese Academy of Sciences, Shanghai 201800, China

**Abstract** By utilizing higher harmonics of undulator radiation, harmonic lasing is helpful in the development of compact x-ray free electron lasers (FELs), i.e. reducing its costs and sizes. Harmonic lasing of FELs have been experimentally demonstrated in the low-gain FEL oscillators from terahertz, infrared to ultraviolet spectral range. Based on the current status and future directions of short-wavelength FELs in the worldwide, this paper reviews the progresses on harmonic lasing of x-ray FELs, mainly concentrating on the recently proposed harmonic lasing of x-ray FEL oscillators and further ideas on harmonic lasing of single pass x-ray FEL amplifiers.

## 1. Introduction

X-rays never stop revolutionizing the understanding of the matters, and creating new sciences and technologies, which have been proven by 20 Nobel Prizes awarded for x-ray related works. Advanced light sources based on particle accelerator, especially synchrotron radiation light source and free electron laser (FEL) hold great prospects as high power, coherent and tunable radiation ranging from the infrared to the hard x-ray regions. Therefore, synchrotron radiation light sources and FELs, devices of relativistic electron beams passing through a periodic magnetic array, are being developed worldwide to satisfy the dramatically growing demands within the material and biological sciences [1].

The successful operation of the first FEL facilities in the XUV and hard x-ray regime [2-5], indicates that the birth of x-ray laser and the era of coherent x-ray science have arrived. Currently, FEL community is on the stage to more sophisticated and determined schemes. On one hand, various ideas are being proposed and studied for improving the x-ray FEL performances of self-amplified spontaneous emission (SASE) [6-7], e.g. in pursuit of temporal coherence [8-14] and fast polarization switch [15-17]. On the other hand, with the rapid progress of the advanced accelerator techniques, scientists begin to envision compact x-ray FEL configurations [18-20] to significantly cut the costs and size of x-ray FEL facilities, which will contribute to the popularization of x-ray FELs, and thus may open up new scientific opportunities.

Utilizing higher harmonics of the undulator radiation is one of the most feasible ways to compact x-ray FEL. The so-called harmonic lasing has been experimentally demonstrated at the long-wavelength regime [21-27], and harmonic lasing schemes of x-ray FEL are under study [18, 28-29]. On the basis of the current status and future directions of x-ray FELs in the worldwide described in Section 2, this paper review the progresses on harmonic lasing of x-ray FELs. We mainly concentrate on the recently proposed harmonic lasing scheme of x-ray FEL oscillator (XFELO) [18, 30] and advanced concepts on harmonic lasing of single-pass x-ray FELs in Section 3 and Section 4, respectively. This paper is concluded with the final remarks in Section5.

## 2. The development of compact x-ray FEL

With the great successful of the linear coherent light source (LCLS) [3], the world's first hard x-ray FEL and Spring-8 angstrom compact free electron laser (SACLA) [4], the world's shortest wavelength x-ray FEL, plenty of x-ray FEL projects are being constructed and planned worldwide, which are driven by the growing interests of FEL users [1].

Up to now, all the hard x-ray FELs use self-amplified spontaneous emission (SASE) as the lasing mode, which starts from the initial shot noise of the electron beam, and results in radiation with excellent spatial coherence, but rather poor temporal coherence. In order to generate fully coherent radiation, various seeded FEL schemes [8-12] were proposed and intensively studied around the world. Echo-enabled harmonic generation (EEHG) [10] is one of the harmonic amplification FEL schemes, which could efficiently work at several tens of harmonic of the seed laser. The successful experimental demonstration of the EEHG mechanism [11] and the first lasing of an EEHG FEL [12] pave the way to coherent soft x-ray radiation from a commercial seed laser. In the hard x-ray regime, the self-seeding approach is the only way that could lead to fully coherent FEL radiation by now. As demonstrated in LCLS, the noisy SASE radiation generated in the first undulator section is spectrally purified by a crystal filter. Then, in the second undulator section, this spectrally purified FEL pulse serves as a highly coherent seed to interact with the electron bunch again to configure a seeded FEL amplifier, which could significantly improve the temporal coherence of the final output [13].

By using an undulator with period length of several centimeters and undulator parameter around unity, a hard x-ray FEL requires electron beam from several to tens of GeV. Thus in order to radiate below 0.15nm, the earliest proposed x-ray FELs are equipped with fairly long linear accelerators, e.g., 14.3GeV room temperature S-band linac for LCLS [3] and 17.4GeV superconductive L-band linac for European XFEL [31]. However, as the second

_________________________

Received 17 Dec 2012

* Supported by Natural Science Foundation of China under Grant No. 11205234 and 11175240

1) E-mail: denghaixiao@sinap.ac.cn

hard x-ray FEL operated in the world, SACLA becomes the first sub-angstrom x-ray FEL at the end of 2011 with 8GeV electron beam and 800m total length [4]. The great successful of SACLA indicates that the high-gradient C-band accelerator is able to work stable for long time, the thermal gun can provide high peak current electron beam with low emittance, and the in-vacuum undulator also serves well enough to make all components working neatly. It is obviously that the large-scale hard x-ray FEL facilities are on their way to miniaturization. Thus, with the rapid progress of accelerator techniques and fruitful experiences from LCLS and SACLA, future x-ray FELs, e.g., the Swiss-FEL (5.8GeV) [32], PAL-XFEL (10GeV) [33] and the compact x-ray FEL proposed in Shanghai (6.3GeV) [34] will be definitely benefited a lot.

In history, the development of the intermediate energy synchrotron radiation light sources has been attributed to high harmonics of the undulator. In comparison with the expensive 6~8GeV large-scale synchrotron light sources, the intermediate synchrotron light sources operating at high harmonics, such as Shanghai synchrotron radiation facility (SSRF) have much lower costs of construction and operation, and are capable of providing comparable performance in 10-25keV photon energy range [35]. The fundamental process of FEL sources involves an electron beam passing through an undulator, which also supports high harmonics with the radiation wavelength inversely scaling with the harmonic order. Thus, high harmonic is an alternative way to offer short-wavelength FEL instead of using high energy electron beam.

Harmonic lasing of the low-gain FEL oscillators was theoretically predicted in early 1980s [21]. The first experimental demonstrations of 3rd harmonic lasing were obtained using infrared FEL oscillators in 1988 [22-23]. Subsequently, the 2nd [24], 3rd and 5th [25] harmonic lasing were carried out at Jefferson infrared oscillator. More recently, the Novosibirsk terahertz FEL oscillator successfully lased on the 3rd harmonic [26] and the NIJI-IV storage ring [27] achieved the 7th harmonic lasing with the optical klystron ETLOK-III. With the development of high-reflectivity high-resolution x-ray crystal [36] and ultra-low emittance electron beams from energy recovering linac (ERL), the low-gain oscillator configuration was reconsidered as a promising candidate for a hard x-ray FEL [30]. Soon afterwards, a harmonic lasing scheme of XFELO was proposed to generate fully coherent x-ray radiation in the spectral of 10-25keV by using a 3.5GeV intermediate energy electron beam [18].

It is widely believed that the high-gain single-pass FEL is the leading candidate in the pursuit of hard x-ray radiation, and harmonic radiations of the high-gain FEL have been theoretically [37] and experimentally [38-39] investigated. However, the harmonic evolution is driven nonlinearly by the fundamental radiation in the high-gain FEL, and thus the harmonic output power. Generally, the saturation power of the 3rd nonlinear harmonic radiation is about 1% of the fundamental. Thus, various harmonic lasing schemes [28-29, 40-43] have been proposed for enhancing the harmonic efficiency in the high-gain FEL from the infrared to hard x-ray spectral region, some of which are of great interest and under considerations for experimental demonstration.

## 3. Harmonic lasing of x-ray FEL oscillator

Generally, the high brightness electron beam from the high coherence mode of an energy recovery linac or the low charge mode of a superconductive linac is utilized for feeding an XFELO. To achieve harmonic lasing in an XFELO [18], the first challenge is the electron beam which can provide sufficient single-pass gain at high harmonics. We assume a 3.5GeV electron beam with the micro-pulse repetition rate of 1MHz, bunch charge of 20pC, normalized emittance of 0.083μm-rad, peak current of 20A and slice energy spread of 100keV. Then an undulator resonant at 3Å, with 15mm period length and 18m total length, shows a 65% single-pass gain at the 3rd harmonic radiation, i.e., 1Å. The second challenge is creating a more favorable lasing condition for the harmonics than that for the fundamental. As shown in Fig. 1, switching the lasing regime from the fundamental to the harmonics can be obtained by a manipulation with the x-ray crystal mirrors, i.e., the Bragg energy $E_H$ of the high selectivity crystal mirrors in XFELO is set to the photon energy of the interested harmonic rather than the fundamental, where $n$ is the interested harmonic order.

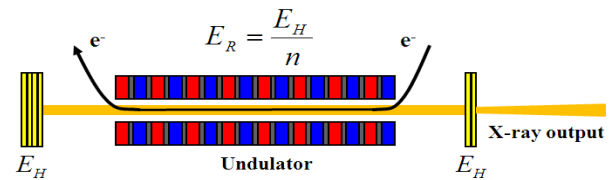

Fig. 1. Harmonic lasing of an x-ray FEL oscillator.

When $(1+g_n) \times (1-c_n) \times r_n > 1$, where $g_n$, $c_n$ and $r_n$ is the single-pass gain, the output coupling efficiency and the round-trip cavity reflectivity of the interested harmonic, respectively, the harmonic radiation evolves from initial spontaneous emission to a coherent pulse, and saturates when $(1+g_n) \times (1-c_n) \times r_n = 1$ because of the gain reduction caused by over-modulation of the strong intra-cavity radiation. The total reflectivity and the output coupling of the x-ray cavity are assumed to be 80% and 5% at the 3rd harmonic, respectively. Then as the peak power growth shown in Fig. 2, an exponential growth of the 3rd harmonic emerges from the initial shot noise start-up after about 30 passes, while the fundamental radiation does not grow significantly.

The single-pass gain sensitivity on the electron beam parameters, i.e., the peak current, transverse emittance and energy spread of the electron beam in the harmonic lasing scheme is more serious than that in the normal XFELO. According to the simulations, a peak current larger than 15A, a normalized transverse emittance less than 0.1μm-rad, and an energy spread less than 150keV should enable a sufficient single-pass gain to start up the 3rd harmonic lasing of an XFELO. In our case, the Rayleigh length of the cavity is 15m, the gain reduction induced by the electron beam imperfection could be compensated by a short Rayleigh range cavity to some extent, which however may degrade the cavity stability.

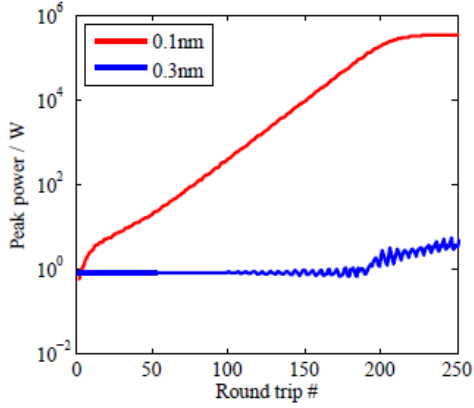

Fig. 2. The 1$^{st}$ and 3$^{rd}$ harmonic peak power growth in a 3$^{rd}$ harmonic lasing XFELO.

For an interested working wavelength, in principle and in simulation, the tolerance requirements on the x-ray cavity in harmonic lasing scheme are similar with that in the fundamental lasing one. Fig. 3 shows the steady-state simulation results of the 3$^{rd}$ harmonic lasing of XFELO, in which the mirror misalignment effects are considered. Since the radiation size on the cavity mirror is 65μm and the mirror aperture diameter is 500μm, a mirror offset of 2μm almost has no influence on the output peak power. The angular tolerance of the mirror may be determined by requiring that the change of the optical axis be much less than the x-ray optical mode angle. Then for a symmetric cavity, the angular tolerance requirement for the mirrors is usually given by

$$\Delta\theta \ll \sqrt{\frac{2\lambda_s}{\pi L_c}}(1-g)^{0.25}(1+g)^{0.75}$$

where $g$, namely the stability parameter of the cavity, is -0.9231 in the discussion. Taking other parameters in the equation, it indicates $\Delta\theta \ll 112$nrad which is consistent with simulation results shown in Fig. 3. It is found that the tolerance requirement on the x-ray cavity alignment is pretty tight but achievable. The further time-dependent simulation shows that, an offset of 2μm combined with an angular tilt of 50nrad is acceptable, which agrees well with the steady-state simulation. Considering there are 2 mirrors at least in an XFELO cavity, the tolerance should be much more stringent.

A lower x-ray saturation power is expected in x-ray cavity for harmonic lasing, corresponding to 3.4μJ pulse energy and 3.4W average power in our case. Therefore, the thermal load effects on the crystals will be much more relaxed, and a linear thermal expansion coefficient less than $1\times10^{-7}$K$^{-1}$ is sufficient to ensure a stable x-ray cavity. In addition, the induced beam energy spread from the FEL interaction in the harmonic lasing x-ray cavity is 350keV, several times smaller than that in the normal XFELO, thus mitigating the beam loss in the ERL return arc and making an ERL operation more straightforward.

Harmonic lasing scheme of XFELO is able to generate MW order coherent x-ray radiation with photon energy of 10-25keV by using a 3.5GeV electron beam. For the 3$^{rd}$ harmonic lasing of XFELO, the numerical example demonstrates a peak brilliance of $1.79\times10^{32}$ photons/(s mm$^2$ mrad$^2$ 0.1%BW) which is comparable with that of a high-gain SASE, e.g. $8.5\times10^{32}$ photons/(s mm$^2$ mrad$^2$ 0.1%BW) for LCLS, and an average brilliance 3 orders of magnitude higher than SASE. With these predicted characters, harmonic lasing of XFELO may contribute to the simplification of x-ray FELs, and may open up new scientific opportunities in various research fields.

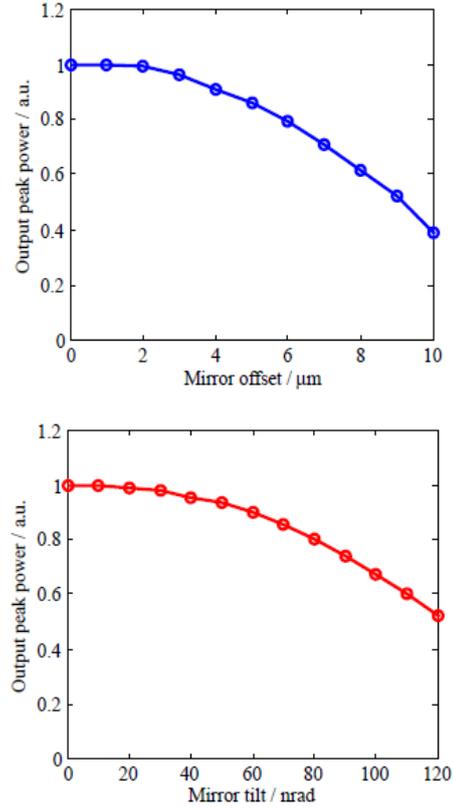

Fig. 3. Performance dependence on the misalignment of crystal mirror for the 3$^{rd}$ harmonic lasing XFELO.

### 4. Harmonic lasing of single-pass x-ray FEL amplifier

In high-gain single-pass FEL, the primary limitation of harmonic lasing is the suppression of the fundamental. As far as our knowledge, Latham proposed [40] the first harmonic lasing scheme of a FEL amplifier in which an injected harmonic signal is amplified to saturation before the fundamental grows up from shot noise. An alternative seeded harmonic lasing scheme of single-pass FEL [41] was suggested by using the same basic elements as high gain harmonic generation (HGHG) [8]. With the chosen parameters of the modulator, the dispersive section, the radiator and the seed laser, the electron density bunching is resonant to a high harmonic of the radiator, instead of the fundamental in standard HGHG. Then the harmonic is rapidly amplified in the radiator while the fundamental radiation of the radiator keeps still at the noise level.

The harmonic growth rate and the harmonic saturation power in the harmonic lasing scheme were intensively studied in refs. [42, 43]. In general, the saturation power of the 3$^{rd}$ harmonic radiation can be enhanced by an order of magnitude in harmonic lasing scheme when compared with nonlinear harmonic generations in single-pass FELs. Recently, because of the success of x-ray FEL operation, various harmonic lasing schemes of single-pass x-ray FEL is under consideration. These proposals mainly play with the phase-shifter between the undulator segments in order to disrupt the electron interaction with the fundamental while the harmonic interaction evolves unhindered, and thus significantly enhance the harmonic radiation efficiency of the existed SASE x-ray FELs.

Coherent x-ray radiation pulse generated from XFELO, atomic x-ray laser [44] and self-seeding SASE FEL [13], can be used as a seed for further FEL amplification. Thus in this section, we propose a seeded harmonic lasing of single-pass FEL to generate fully coherent x-ray pulses. The key point of this proposal is using an ultimate storage ring driven XFELO builds fully coherent x-ray radiation at a relatively longer-wavelength in the Bragg cavity firstly, as shown in Fig. 4, then the intra-cavity radiation is used to modulate the 3.5GeV electron beam from a high brightness linac, and serves as a seed laser of the seeded harmonic lasing scheme for generating high harmonic radiation in the final radiator.

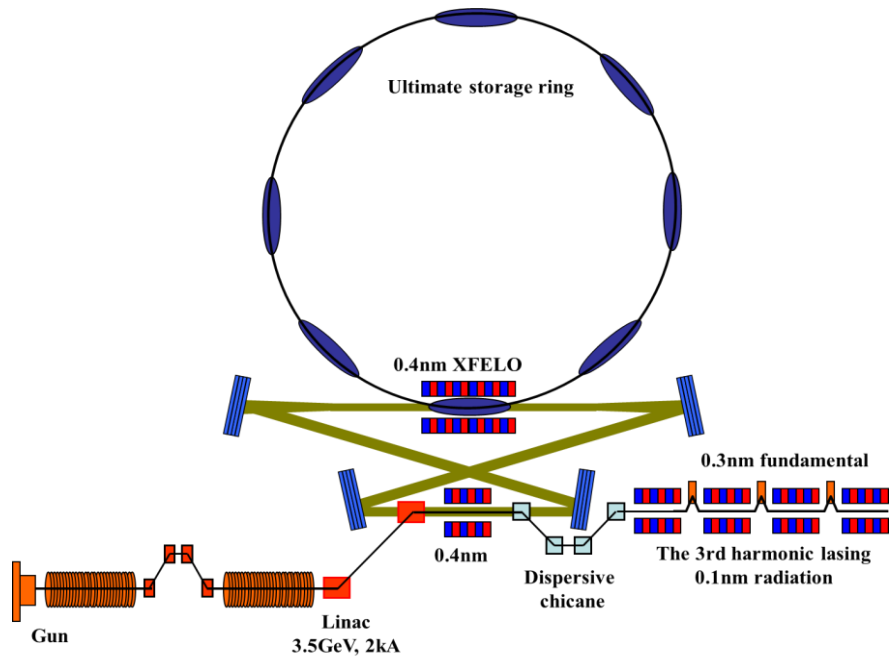

Fig. 4. Schematic of seeded harmonic lasing scheme of a single-pass x-ray FEL.

Ultimate storage ring is widely considered to be next generation of synchrotron radiation light source, since it may provide electron beam with pretty small transverse emittance which is important to synchrotron light source. Generally, a high-gain x-ray FEL is not possible to be operated on an ultimate storage ring because of the large energy spread of the electron beam. However, a low-gain FEL oscillator is definitely suitable for ultimate storage ring, considering that the existed storage ring based FEL oscillators in operation. If one taking PEP-X parameters [45] as an example, i.e., 4.5GeV beam energy, $1.3 \times 10^{-3}$ beam energy spread, 160pm/80pm geometrical emittance and 300A peak current, a single-pass gain above 50% can be modeled for 0.4nm radiation within a 20m long undulator. It is large enough to start up the power growth and store a strong intra-cavity radiation. Therefore, an XFELO resonant at 0.4-1nm can be established on an ultimate storage ring.

Considering the tendency that synchrotron light source and linac based x-ray FEL will be placed in one campus, we investigate a case where the electron beams from the storage ring and the linac are synchronized. The 4-5GeV ultimate storage ring is used to drive a normal x-ray FEL oscillator which is resonant at 0.4nm. By choosing the thickness of the Bragg crystal, the output fraction and hence the saturation power in the 0.4nm cavity can be changed. The powerful radiation field in the cavity is utilized as a seed laser for a linac-based high-gain FEL. The electron beam is supposed to be generated from the linac low-charge mode, with 3.5GeV beam energy, 2kA peak current, 300keV sliced beam energy spread and 0.083μmrad normalized transverse emittance. After the modulator and the dispersive chicane, the electron beam enters into the radiator with abundant harmonic bunching of 0.4nm. With the proper choice of undulator parameter, the 4$^{th}$ harmonic of the seed laser is the 3$^{rd}$ harmonic of the radiator, and then the 0.1nm radiation will be rapidly produced and amplified exponentially until saturation. At the same time, the 0.3nm fundamental radiation of the radiator is expected to start from shot noise.

Fig. 5 shows the simulated peak power growth of the 3$^{rd}$ harmonic and the fundamental in a harmonic lasing scheme shown in Fig. 4. Finally, the 0.1nm radiation achieves a peak power larger than 1GW, while the peak power of the 0.3nm radiation is still 4MW. With a peak power of GW order, the peak brilliance of this scheme will be 3-4 orders higher than a harmonic lasing scheme of an XFELO and current SASE x-ray FEL. Moreover, it opens an opportunity for the pump-probe manipulation between the synchrotron light source and x-ray FEL.

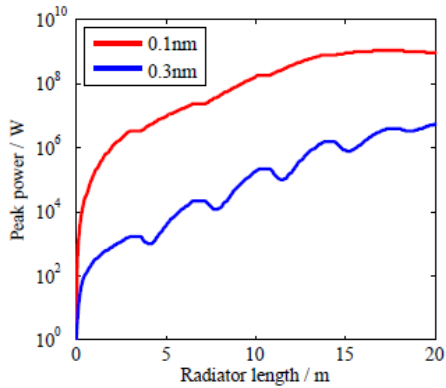

Fig. 5. Simulated peak power evolution in the harmonic lasing of single-pass x-ray FEL.

It is worth stressing that an ERL can easily substitute the ultimate storage ring here. However, when compared with an ultimate ring, the disadvantage arisen from an ERL is the substantive superconductive linac modules. Furthermore, the seeded harmonic lasing scheme is well suited for the recently proposed concept where the x-ray FEL and ERL share a common superconducting electron linac simultaneously [46], then the beam energy of the superconducting linac can be reduced from 5-7GeV to 3-4GeV for a specifically interested wavelength, or the radiation wavelength may cover even shorter range with the fixed electron beam energy.

## 5. Conclusions

In conclusion, compared with the conventional FEL operating at the fundamental frequency, harmonic lasing scheme of x-ray FEL produces significant power using an electron beam with intermediate energy, e.g., 3.5GeV, which theoretically offers a way to smaller and cheaper x-ray FEL. In this paper, harmonic lasing of x-ray FEL oscillator and harmonic lasing of single-pass x-ray FEL amplifier are studied. These compact x-ray FEL schemes could be one of the candidates for the x-ray FEL, ERL and ultimate ring light source in the near future.

Besides the contribution to the compact of x-ray FEL, harmonic lasing is of great interest and importance more generally. On one hand, harmonic lasing of an existing short undulator can be utilized to characterize the high harmonic bunching in a seeded FEL [47], which presents a better resolution and sensitivity of the diagnoses in comparison with the coherent transition radiation based method. On other hand, harmonic lasing was proposed to improve the temporal coherence and spectral brightness of a SASE FEL, where a few undulator segments in the middle stage of the exponential growth regime operates at harmonic lasing mode, thus effectively reduce the FEL bandwidth through a slippage-boosted way [48].

*The authors would like to thank Alexander Chao from SLAC, Senyu Chen from IHEP, Tong Zhang, Chao Feng, Dong Wang and Zhentang Zhao from SINAP for helpful discussions.*